\documentclass[usenatbib,useAMS]{mn2e}

\usepackage{graphicx}

\newcommand{\Msun}{\ensuremath{M_{\odot}}}
\newcommand{\MgFe}{\ensuremath{[{\rm MgFe}]^{\prime}}}
\newcommand{\Hb}{\ensuremath{{\rm H}\beta}}
\newcommand{\Mgtwo}{\ensuremath{{\rm Mg}_2}}
\newcommand{\Mgb}{\ensuremath{{\rm Mg}\, b}}
\newcommand{\Fe}{\ensuremath{\langle {\rm Fe}\rangle}}
\newcommand{\aFe}{\ensuremath{\alpha/{\rm Fe}}}

\newcommand{\CaTs}{\ensuremath{{\rm CaT}^*}}
\newcommand{\aCa}{\ensuremath{\alpha/{\rm Ca}}}
\newcommand{\FeH}{\ensuremath{{\rm Fe}/{\rm H}}}
\newcommand{\ZH}{\ensuremath{Z/{\rm H}}}
\newcommand{\CNone}{\ensuremath{{\rm CN}_1}}
\newcommand{\TiOtwo}{\ensuremath{{\rm TiO}_2}}

\title[New Clues on the Calcium Underabundance in Early-Type Galaxies]
{New Clues on the Calcium Underabundance in Early-Type Galaxies}

\author[Daniel Thomas, Claudia Maraston, \& Ralf Bender] {Daniel
Thomas$^1$, Claudia Maraston$^1$, \& Ralf Bender$^{1,2}$\\
$^1$Max-Planck-Institut f\"ur extraterrestrische Physik,
Giessenbachstra\ss e, D-85748 Garching, Germany\\
$^2$Universit\"ats-Sternwarte M\"unchen, Scheinerstr.~1, D-81679
M\"unchen, Germany}

\date{Accepted 2003 March 27. Received {\ldots} ;
      in original form 2003 March 11}

\pagerange{\pageref{firstpage}--\pageref{lastpage}}

\pubyear{2003}

\begin{document}

\maketitle

\label{firstpage}

\begin{abstract}
We use our new stellar population models, which include effects from
variable element abundance ratios, to model the Ca4227 absorption line
indices of early-type galaxies \citep{Traetal98}, and to derive
calcium element abundances. We find that calcium, although being an
$\alpha$-element, is depressed with respect to the other
$\alpha$-elements by up to a factor 2. This confirms quantitatively
earlier speculations that early-type galaxies are calcium
underabundant. We find a clear correlation between \aCa\ ratio and
central velocity dispersion, which implies that more massive galaxies
are more calcium underabundant. Interestingly this correlation extends
down to the dwarf spheroidal galaxies of the Local Group for which
\aCa\ ratios have been measured from high-resolution spectroscopy of
individual stars (Shetrone et al.).  The increase of the calcium
underabundance with galaxy mass balances the higher total
metallicities of more massive galaxies, so that calcium abundance in
early-type galaxies is fairly constant and in particular does not
increase with increasing galaxy mass.  This result may be the key to
understand why the CaII triplet absorption of early-type galaxies at
$8600$~\AA\ is constant to within 5 per cent over a large range of
velocity dispersions \citep{Sagetal02,Cenetal03}. The origin of the
calcium underabundance in early-type galaxies remains yet to be
understood. We argue that formation timescales are disfavoured to
produce calcium underabundance, and that the option of metallicity
dependent supernova yields may be the most promising track to follow.
\end{abstract}


%================= I N T R O D U C T I O N ===============================

\section{Introduction}
\label{sec:intro}
In the late seventies it has first been noticed that early-type
galaxies show relatively weak CaII triplet (CaT) absorption at
8600~\AA\ \citep{Cohen79,FF80}. The metallicities derived from these
lines lie well below solar, which seems hard to reconcile with the
super-solar metallicities obtained from the classical Lick indices
\Mgtwo\ and \Fe\
\citep[e.g.,][]{Traetal00a,Kunetal01,TMB02b,Maretal03}.  Moreover, the
very tight correlation between \Mgtwo\ and velocity dispersion
$\sigma$ \citep*[e.g.,][]{BBF93} has not been found for the CaT line
\citep{Cohen79,FF80}.  Subsequent measurements of CaT absorption in
early-type galaxies providing increasingly better data quality confirm
these early results (\citealt*{TDT90}; \citealt{Peletal99}). Very
recently, \citet{Sagetal02} measured the CaT absorption in 94
early-type galaxies at very high signal-to-noise ratios, using an
improved index definition \citep{Cenetal01}. They again find a
relatively weak CaT absorption and even a mild anti-correlation
between CaT and $\sigma$ with a remarkably small scatter \citep[see
also][]{Cenetal03}.

\citet{Sagetal02} show that neither uncertainties in the CaT modelling
nor an (already previously suggested) steepening of the initial mass
function provide compelling explanations. A very attractive way out
remains to be the conclusion that early-type galaxies are indeed
underabundant in the element calcium. The main problem is that this
interpretation remains speculative as long as the sensitivity of the
CaT index to calcium abundance is not known, and in particular calcium
underabundant stellar population models of CaT are not available.

Interestingly, also the blue Lick index Ca4227 at $\lambda\approx
4227$~\AA\ is very weak in early-type galaxies, which again indicates
a possible calcium underabundance
\citep{Worthey92,Vazetal97,Traetal98}. A similar conclusion has been
drawn on the basis of Ca4455 index measurements
\citep{Worthey98,HW99}. We note, however, that Ca4455, despite its
name, appears to be only sensitive to the abundances of iron-peak
elements \citep{TB95}, so that the weak Ca4455 absorption reflects the
well-known iron underabundance in early-type galaxies
\citep*[e.g.,][]{WFG92}. Ca4227, instead, is sensitive to calcium
abundance, and---different from CaT---for Ca4227 we now have a set of
stellar population models at hand \citep*[][hereafter TMB]{TMB03} that
explicitly take abundance effects of individual elements into account,
and in particular are computed for different calcium abundances. Here
we use these models to analyse the data of \citet{Traetal98}, and for
the first time to quantitatively derive calcium element abundances of
early-type galaxies.

In Section~2 we briefly introduce the stellar population model. The
results are presented in Section~3 and discussed in Section~4.

%========================= M O D E L =====================================

\section{Models}
\label{sec:construction}
In TMB we present stellar population models for the first time with
different chemical mixtures and element abundance ratios. We computed
all optical Lick indices from \CNone\ to \TiOtwo\ in the wavelength
range $4000\la\lambda\la 6500$~\AA\ of Simple Stellar Population (SSP)
models varying the abundance ratios of $\alpha$-elements (i.e.\ O, Mg,
Ca, Na, Si, Ti) to iron peak elements (Fe, Cr) and to the individual
elements carbon, nitrogen, and calcium.  The impact from the element
abundance changes on the Lick absorption-line indices are taken from
\citet{TB95}.  The models comprise ages between $1$ and $15$~Gyr and
metallicities between 1/200 and 5 solar. We refer to TMB for more
details.

Particular care was taken to calibrate the SSP models on globular
cluster data, which---most importantly---include objects with
relatively high, namely solar, metallicities \citep{Puzetal02}. We
match very well their Ca4227 indices with models in which calcium is
enhanced like the other $\alpha$-elements with respect to the
iron-peak elements, in agreement with results from high-resolution
spectroscopy of individual stars (see TMB). Note that calcium itself
is an $\alpha$-element. Having the issue of weak calcium absorption in
early-type galaxies in mind, we developed additional models in which
calcium is detached from the group of $\alpha$-elements, allowing for
variations of the \aCa\ element ratio.  In TMB, models of the index
Ca4227 for the element ratios $[\aCa]=-0.1,\:0.0,\:0.2,\:0.5$ are
provided, with $[\aCa]\equiv \log \left({X_{ \alpha}}/{X_{\rm
Ca}}\right) - \log \left({X_{\alpha}}/{X_{\rm Ca}}\right)_{\odot}$.

We will use these models to derive \aCa\ ratios of early-type
galaxies.  Note that the above notation relates the element ratio to
the solar value. We will speak of calcium underabundance, if \aCa\ is
larger than solar, hence $[\aCa]>0$.

\section{Results}
We analyse the early-type galaxy data of \citet{Traetal98}. This
turned out to be the only data set in the literature with Ca4227
indices for 'normal' early-type galaxies. We do not use the sample of
\citet{Lonetal00}, because it consists of peculiar, disturbed and
shell galaxies. Still, it should be kept in mind that the
\citet{Traetal98} sample contains a rather inhomogeneous morphological
mix, with many S0 and later-type galaxies included.  From this sample
we selected those objects, for which the error in the Ca4227
measurement is smaller than 0.25~\AA, which yields 39 galaxies with a
median error of 0.2~\AA.

\subsection{\boldmath\MgFe\ vs.\ Ca4227}
\begin{figure*}
\includegraphics[width=\linewidth]{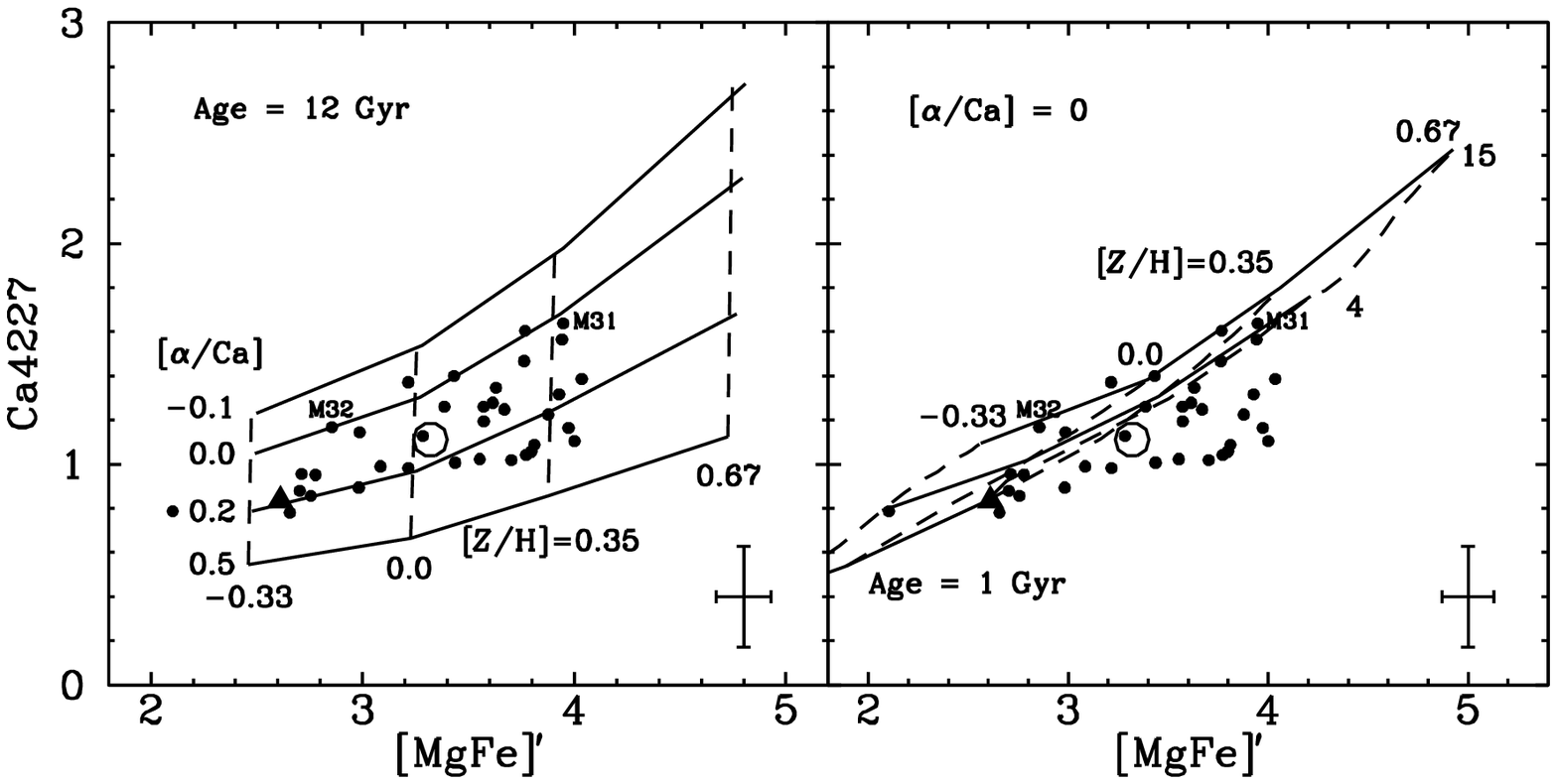}
\caption{Early-type galaxies in the \MgFe-Ca4227 plane.  Galaxy data
(filled circles) are from \citet{Traetal98}. The open circle is their
median. The filled triangle is the integrated Galactic Bulge light
from \citet{Puzetal02}. Lines are stellar population models from
\citet{TMB03}. {\em Left-hand panel:} Models for constant \aCa\ ratios
$[\aCa]=-0.1,\:0.0,\:0.2,\:0.5$ (solid lines) and constant
metallicities $[\ZH]=-0.33,\ 0.0,\ 0.35,\ 0.67$ (dashed lines) at a
fixed age $t=12$~Gyr. {\em Right-hand panel:} Models for constant ages
of 1, 4, and 15~Gyr (solid lines), and the constant metallicities
$[\ZH]=-0.33,\ 0.0,\ 0.35,\ 0.67$ (dashed lines) at a fixed solar
\aCa\ element ratio ($[\aCa]=0$).}
\label{fig:indices}
\end{figure*}
In Fig.~\ref{fig:indices} we show Ca4227 as a function of the index
\MgFe, with $\MgFe\equiv \sqrt{\Mgb\cdot (0.72\cdot {\rm
Fe5270}+0.28\cdot {\rm Fe5335})}$ as defined by TMB. \MgFe\ is
independent of \aFe\ ratio variations, and therefore a good tracer of
total metallicity. SSP models for the \aCa\ ratios
$[\aCa]=-0.1,\:0.0,\:0.2,\:0.5$ and the metallicities
$[\ZH]=-0.33,\:0.0,\:0.35,\:0.67$ at the fixed age $t=12$~Gyr are
plotted in the left-hand panel (see labels in the figure).  Models of
constant metallicity and varying \aCa\ ratio (dashed lines) are almost
perfectly vertical, indicating that \MgFe\ is not sensitive to calcium
abundance. Solid lines in the left-hand panel of
Fig.~\ref{fig:indices} are models with constant \aCa\ ratios. Note
that the models in the \MgFe-Ca4227 plane are highly degenerate in
age, as both indices have similar age dependencies. This is shown with
the right-hand panel of Fig.~\ref{fig:indices}, in which models with
ages 1, 4, and 15~Gyr and metallicities
$[\ZH]=-0.33,\:0.0,\:0.35,\:0.67$ are plotted (see labels in the
figure), the \aCa\ ratio being fixed at the solar value. The left-hand
panel of Fig.~\ref{fig:indices} therefore is well suited for isolating
\aCa\ ratio effects.

The galaxy data are shown as filled circles, the open circle is their
median. The filled triangle is the integrated light of the Galactic
Bulge \citep{Puzetal02}.  The data lie below the model with solar
\aCa\ ratio, no combination of age and metallicity can be found to
match the data.  \aCa\ ratios from solar to roughly 3 solar, $0\la
[\aCa]\la 0.5$, are obtained.  The maximal calcium underabundance
relative to the other $\alpha$-elements thus is about a factor
$2\,$--$\,3$. We are confident that this detection of calcium
underabundance is significant in spite of the relatively large
observational errors for two reasons.  1) The Ca4227 range covered by
the data plotted here is consistent with the data of
\citet{Lonetal00}, which have, although being disturbed and shell
galaxies, smaller observational errors ($<0.09$~\AA).  2) The median
of the sample (open circle) is at ${\rm Ca4227}=1.1\pm 0.03$~\AA,
significantly below the model with solar \aCa. From the location of
the median we infer that calcium is depleted in early-type galaxies by
typically a factor 1.4.

The large triangle in Fig.~\ref{fig:indices} is the average of 15
Galactic Bulge fields measured by \citet{Puzetal02}.  Obviously, the
Galactic Bulge is also calcium underabundant. Its \aCa\ ratio is
consistent with the median of the early-type galaxy sample, at
somewhat lower total metallicity.  This result is in good agreement
with the detailed abundance analysis of individual field Bulge stars
from high-resolution spectra by \citet{McWR94}. These authors measure
super-solar \aCa\ ratios ($[\aCa]\sim 0.2$), hence calcium
underabundance, for the metal-rich stars ($[\FeH]>-0.5$) of their
sample.

The galaxies M31 and M32 are labelled in Fig.~\ref{fig:indices}. M31
has stronger Ca4227 and \MgFe\ indices than M32, so that both objects
have very similar \aCa\ ratios ($[\aCa]\sim 0$). Curiously, M32 shows
stronger CaT absorption than M31 \citep{KB99}, which seems to indicate
higher calcium abundance in M32. Note however, that the CaT index is
severely contaminated by Paschen absorption lines
\citep{Cenetal01}. As M32 is younger than M31 \citep{Traetal00b}, the
strong CaT absorption index of M32 is most likely caused by an
increase of Paschen line absorption.

\subsection{Correlations with galaxy velocity dispersion}
\begin{figure}
\includegraphics[width=\linewidth]{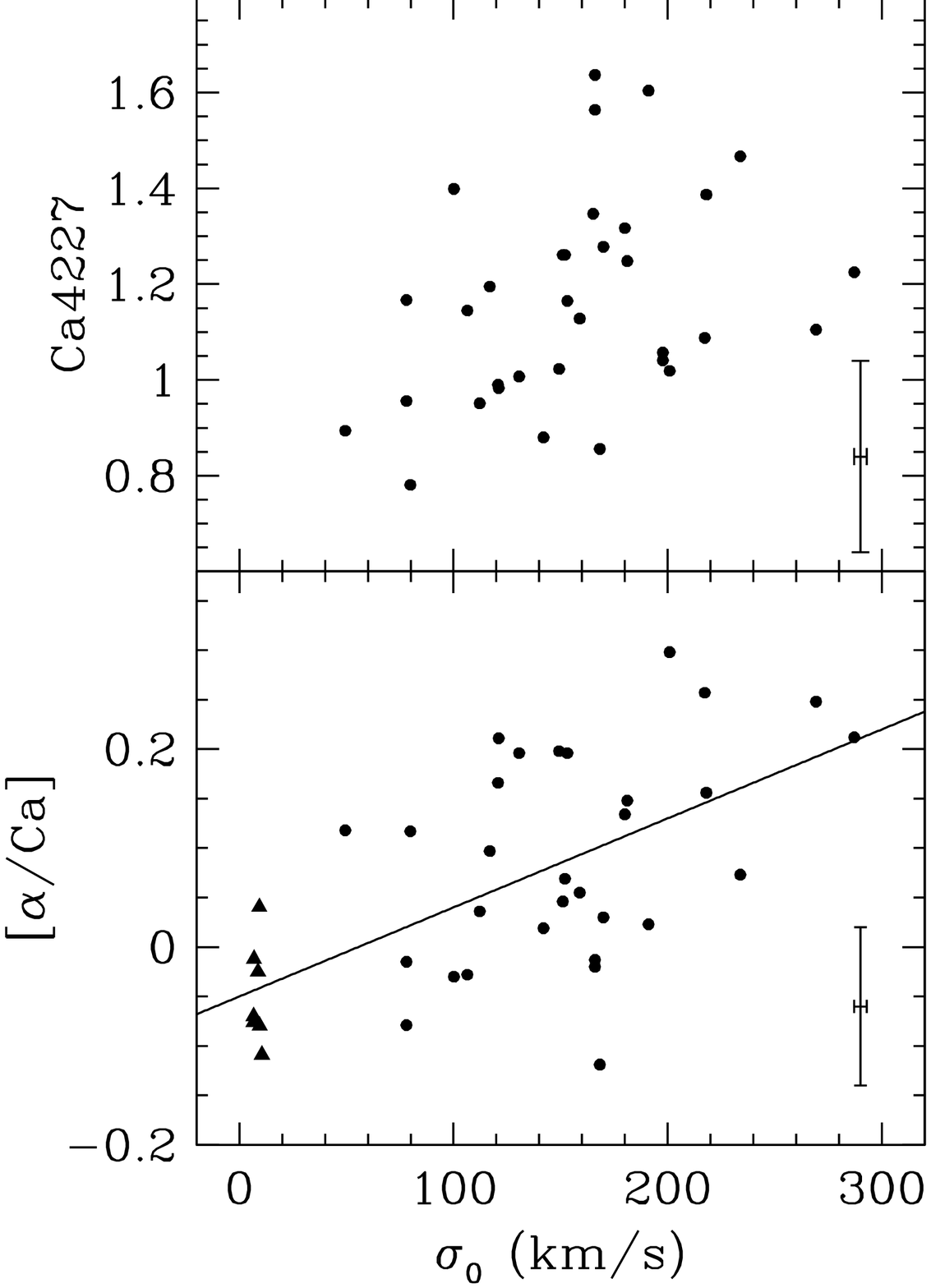}
\caption{Ca4227 index (top panel) and \aCa\ ratio (bottom panel) of
early-type galaxies as functions of central velocity dispersion.
Ca4227 data are from \citet[][circles]{Traetal98}. \aCa\ ratios are
derived from Fig.~\ref{fig:indices}. Triangles are Local Group dwarf
spheroidal galaxies for which \aCa\ ratios are measured in individual
stars \citep{SCS01,Sheetal03}. Central velocity dispersions are from
\citet{WMT85}, \citet{Fabetal89}, \citet*{BBF92}, \citet{G93}, and
\citet{Mateo98}. The solid line is a linear least-square fit to the
giant galaxy data (circles).}
\label{fig:relation}
\end{figure}
In Fig.~\ref{fig:relation}, the index Ca4227 (top panel) and the \aCa\
ratios derived from Fig.~\ref{fig:indices} (bottom panel) are plotted
as functions of central velocity dispersion (circles). More precisely,
we determine with the TMB models ages, metallicities, and \aCa\ ratios
from the indices \Hb, \MgFe, and Ca4227. It should be emphasized, that
uncertainties in the stellar population ages have only a minor effect
on the determination of the \aCa\ ratio as demonstrated by
Fig.~\ref{fig:indices}.  Velocity dispersions are taken from
\citet*{WMT85}, \citet{Fabetal89}, \citet{BBF92}, and \citet{G93}.
The triangles in Fig.~\ref{fig:relation} are the Local Group dwarf
spheroidal galaxies Carina, Draco, Fornax, Leo I, Sextans, and Ursa
Minor. Their mean \aCa\ ratios are the average of the element
abundances of individual stars ($\sim 5$ per object) determined from
high-resolution spectroscopy \citep{SCS01,Sheetal03}. These authors
derive $\alpha$-element abundances by averaging the abundances of the
elements Mg, Ca, and Ti. Central velocity dispersions are taken from
the compilation of \citet{Mateo98}.

Ca4227 correlates with velocity dispersion within the measurement
errors. The increase of Ca4227 with increasing $\sigma_0$, however, is
balanced by the simultaneous increase of \MgFe, so that also \aCa\
increases with velocity dispersion. A linear least-square fit (solid
line) to the giant galaxy data yields
\begin{equation}
[\aCa]= -0.03 + 0.0008*\sigma_0\ .
\end{equation}
In other words, more massive galaxies are more metal-rich and more
calcium underabundant.  The admittedly large scatter of the relation
between \aCa\ and $\sigma_0$ is fully consistent with the errors in
[\aCa] ($\sim 0.1$~dex), which are caused mainly by the large
observational errors of Ca4227. Ca4227 is a rather weak absorption
line, and requires high signal-to-noise to be measured accurately.
Given these large errors, it seems even surprising that such a clear
correlation could be found. Most interestingly, it is very well
consistent with the \aCa\ ratios of the Local Group dwarf spheroidals.
We note that the large spread in Fig.~\ref{fig:relation} may also come
from the inhomogeneous morphological mix in the \citet{Traetal98}
sample.  Ca4227 index measurements for a similarly large sample of
elliptical galaxies at significantly higher signal-to-noise than the
\citet{Traetal98} data would be very useful to check the present
result.

The increase of [\aCa] by $\sim 0.2$ dex between $\sigma_0=100$ and
300~km/s is comparable to the typical metallicity increase
\citep[][D.\ Thomas et al., in preparation]{Traetal00b,Kunetal01} in
that mass range. Hence calcium abundance stays approximately constant
or may even decrease (the present data quality does not allow us to
discriminate these possibilities). The Ca4227 index increases, mainly
because of its sensitivity not only to calcium abundance but also to
total metallicity \citep{TB95}. The Paschen line corrected CaII
triplet index \CaTs\ at 8600~\AA\ \citep{Cenetal01}, instead, is found
to be constant to within 5 per cent over the same range of velocity
dispersions \citep{Sagetal02,Cenetal03}. This result can be understood
now, provided that the index \CaTs\ is more sensitive to calcium
abundance than Ca4227.

\section{Discussion and Conclusions}
In the previous section we show that early-type galaxies are calcium
underabundant.  The reason for this chemical peculiarity is not yet
understood.  In this section we will briefly discuss possible
mechanisms that can lead to the formation of calcium underabundant
stellar populations.

\subsection{Formation timescales and IMF}
More massive early-type galaxies have higher \aFe\ ratios and older
average ages \citep[e.g.,][]{Traetal00b,TMB02b,TF02,CRC03}, which
indicates faster formation histories of their stellar populations
\citep*{Ma94,TGB99}. Note that high \aFe\ ratio are actually an iron
underabundance \citep[][TMB]{BMG94,Traetal00a}. They are a good
measure for star formation timescales, because a substantial fraction
of iron comes from the delayed enrichment by Type Ia supernovae
\citep*{GR83,MG86,TGB98}. Note that also the old, metal-poor stars in
the halo of the Milky Way, which formed at an early stage of the
Galaxy's evolution, are underabundant in iron \citep[][and references
therein]{McW97}. It may seem therefore straightforward to link also
the underabundance of calcium with formation timescales.

The situation for the \aCa\ ratio, however, is entirely different. The
old, metal-poor halo stars are not underabundant in calcium
\citep{McW97}, hence the solar \aCa\ ratio is already at place in the
very beginning of chemical enrichment. In other words, calcium, which
is actually an $\alpha$-element, is enriched in lockstep with the
other $\alpha$-elements. Moreover, Type Ia supernovae produce mainly
iron and do not contribute significantly to the enrichment of calcium
\citep*{NTY84}. The calcium underabundance found in early-type
galaxies can therefore not be explained through the link of formation
timescales with delayed Type Ia supernova enrichment.

Within Type~II supernova nucleosynthesis, calcium comes mainly from
supernovae of the less massive progenitors with $m\la 20$~\Msun\
\citep*{WW95,TNH96}. If the chemical enrichment is dominated by stars
more massive than $\sim 20~\Msun$, calcium underabundance can be
achieved \citep{MG00}. Because of the short lifetimes of these stars,
however, extremely short formation timescales ($\sim 10^7$~yr) would
be required, implying unreasonably high star formation rates up to
$10^5$~\Msun\ per year.

Alternatively, also a flattening of the initial mass function (IMF)
enhances the contribution from high-mass stars to the chemical
enrichment. Integrating the supernova yields of \citet{WW95} with
variable IMF slopes, we find that an abundance ratio $[\aCa]\sim 0.2$
requires extremely shallow IMF slopes of the order $x\sim 0.2$
(Salpeter being $x=1.35$). A flattening of the initial mass function
to produce calcium underabundance can therefore be ruled out, as such
shallow IMFs violate other observational constraints, in particular
predict stellar mass-to-light ratios \citep{Ma98} well above the
dynamical determinations \citep{Geretal01}.

\subsection{Metallicity-dependent yields}
A more appealing solution may be connected to a metallicity dependence
of supernova yields, in the sense that in metal-rich supernovae the
production (or at least the ejection) of calcium is depressed
\citep{Worthey98}.

From the theoretical side of Type~II supernova modelling, there are no
clear indications for the existence of metallicity-dependent yields.
The calculations of \citet[][model B]{WW95} predict only a very mild
increase of the \aCa\ ratio with increasing metallicity in the
supernova ejecta for the metallicity range 1/10,000 to solar.
Extrapolating these yields to double-solar metallicity leads to [\aCa]
ratios, which are $\sim 0.2$ dex below the values derived here for
massive early-type galaxies. If metallicity effects are at work, they
must set in at super-solar metallicities, a range which has not yet
been explored in Type~II supernova nucleosynthesis calculations.

This option gets support from the fact that only the stars with
metallicities above solar in the \citet{McWR94} data are calcium
underabundant. It would also be very well consistent with the
\aCa-$\sigma$ correlation of early-type galaxies found in this
paper. The most significant increase of the \aCa\ ratio with galaxy
mass occurs among giant galaxies in a regime of super-solar
metallicity.  The decrease of the \aCa\ ratio when going to the Local
Group dwarf spheroidals by at most 0.1~dex is relatively small,
instead, given the huge drop in metallicity to about 1.5~dex below
solar. To conclude, the possibility of metallicity dependent yields at
super-solar metallicities remains a promising track to follow in order
to find the origin of the calcium underabundance in early-type
galaxies.

\section*{Acknowledgments}
We would like to thank Scott Trager for the many interesting and
fruitful discussions.  We also acknowledge the anonymous referee for
the very constructive comments on the manuscript, in particular for
drawing our attention on the dwarf spheroidal data.

%========================= R E F E R E N C E S ===========================

%
%
%========================= R E S U L T S =================================
%
\label{lastpage}
\bsp


\small
\begin{thebibliography}{}

\bibitem[\protect\citeauthoryear{Bender, Burstein \& Faber}{Bender
  et~al.}{1992}]{BBF92}
Bender R.,  Burstein D.,    Faber S.~M.,  1992, ApJ, 399, 462

\bibitem[\protect\citeauthoryear{Bender, Burstein \& Faber}{Bender
  et~al.}{1993}]{BBF93}
Bender R.,  Burstein D.,    Faber S.~M.,  1993, ApJ, 411, 153

\bibitem[\protect\citeauthoryear{Buzzoni, Mantegazza \& Gariboldi}{Buzzoni
  et~al.}{1994}]{BMG94}
Buzzoni A.,  Mantegazza L.,    Gariboldi G.,  1994, AJ, 107, 513

\bibitem[\protect\citeauthoryear{Caldwell, Rose \& Concannon}{Caldwell
  et~al.}{2003}]{CRC03}
Caldwell N.,  Rose J.~A.,    Concannon K.~D.,  2003, AJ, in press,
  astro-ph/0303345

\bibitem[\protect\citeauthoryear{Cenarro, Cardiel, Gorgas, Peletier, Vazdekis
  \& Prada}{Cenarro et~al.}{2001}]{Cenetal01}
Cenarro A.~J.,  Cardiel N.,  Gorgas J.,  Peletier R.~F.,  Vazdekis A.,    Prada
  F.,  2001, MNRAS, 326, 959

\bibitem[\protect\citeauthoryear{{Cenarro}, {Gorgas}, {Vazdekis}, {Cardiel} \&
  {Peletier}}{{Cenarro} et~al.}{2003}]{Cenetal03}
{Cenarro} A.~J.,  {Gorgas} J.,  {Vazdekis} A.,  {Cardiel} N.,    {Peletier}
  R.~F.,  2003, MNRAS, 339, L12

\bibitem[\protect\citeauthoryear{Cohen}{Cohen}{1979}]{Cohen79}
Cohen J.~G.,  1979, ApJ, 228, 405

\bibitem[\protect\citeauthoryear{Faber \& French}{Faber \& French}{1980}]{FF80}
Faber S.~M.,  French H.~B.,  1980, ApJ, 235, 405

\bibitem[\protect\citeauthoryear{Faber, Wegner, Burstein, Davies, Dressler,
  Lynden-Bell \& Terlevich}{Faber et~al.}{1989}]{Fabetal89}
Faber S.~M.,  Wegner G.,  Burstein D.,  Davies R.~L.,  Dressler A.,
  Lynden-Bell D.,    Terlevich R.~J.,  1989, ApJS, 69, 763

\bibitem[\protect\citeauthoryear{Gerhard, Kronawitter, Saglia \&
  Bender}{Gerhard et~al.}{2001}]{Geretal01}
Gerhard O.,  Kronawitter A.,  Saglia R.~P.,    Bender R.,  2001, AJ, 121, 1936

\bibitem[\protect\citeauthoryear{Gonz{\'{a}}lez}{Gonz{\'{a}}lez}{1993}]{G93}
Gonz{\'{a}}lez J.,  1993, Phd~thesis, University of California, Santa Cruz

\bibitem[\protect\citeauthoryear{Greggio \& Renzini}{Greggio \&
  Renzini}{1983}]{GR83}
Greggio L.,  Renzini A.,  1983, A\&A, 118, 217

\bibitem[\protect\citeauthoryear{{Henry} \& {Worthey}}{{Henry} \&
  {Worthey}}{1999}]{HW99}
{Henry} R.~B.~C.,  {Worthey} G.,  1999, PASP, 111, 919

\bibitem[\protect\citeauthoryear{Kormendy \& Bender}{Kormendy \&
  Bender}{1999}]{KB99}
Kormendy J.,  Bender R.,  1999, ApJ, 522, 772

\bibitem[\protect\citeauthoryear{Kuntschner, Lucey, Smith, Hudson \&
  Davies}{Kuntschner et~al.}{2001}]{Kunetal01}
Kuntschner H.,  Lucey J.~R.,  Smith R.~J.,  Hudson M.~J.,    Davies R.~L.,
  2001, MNRAS, 323, 615

\bibitem[\protect\citeauthoryear{Longhetti, Bressan, Chiosi \&
  Rampazzo}{Longhetti et~al.}{2000}]{Lonetal00}
Longhetti M.,  Bressan A.,  Chiosi C.,    Rampazzo R.,  2000, A\&A, 353, 917

\bibitem[\protect\citeauthoryear{Matteucci}{Matteucci}{1994}]{Ma94}
Matteucci F.,  1994, A\&A, 288, 57

\bibitem[\protect\citeauthoryear{McWilliam}{McWilliam}{1997}]{McW97}
McWilliam A.,  1997, ARA\&A, 35, 503

\bibitem[\protect\citeauthoryear{McWilliam \& Rich}{McWilliam \&
  Rich}{1994}]{McWR94}
McWilliam A.,  Rich R.~M.,  1994, ApJS, 91, 749

\bibitem[\protect\citeauthoryear{Maraston}{Maraston}{1998}]{Ma98}
Maraston C.,  1998, MNRAS, 300, 872

\bibitem[\protect\citeauthoryear{Maraston, Greggio, Renzini, Ortolani, Saglia,
  Puzia \& Kissler-Patig}{Maraston et~al.}{2003}]{Maretal03}
Maraston C.,  Greggio L.,  Renzini A.,  Ortolani S.,  Saglia R.~P.,  Puzia T.,
    Kissler-Patig M.,  2003, A\&A, 400, 823

\bibitem[\protect\citeauthoryear{Mateo}{Mateo}{1998}]{Mateo98}
Mateo M.~L.,  1998, ARA\&A, 36, 435

\bibitem[\protect\citeauthoryear{Matteucci \& Greggio}{Matteucci \&
  Greggio}{1986}]{MG86}
Matteucci F.,  Greggio L.,  1986, A\&A, 154, 279

\bibitem[\protect\citeauthoryear{Moll{\'a} \& Garc{\'\i}a-Vargas}{Moll{\'a} \&
  Garc{\'\i}a-Vargas}{2000}]{MG00}
Moll{\'a} M.,  Garc{\'\i}a-Vargas M.~L.,  2000, A\&A, 359, 18

\bibitem[\protect\citeauthoryear{Nomoto, Thielemann \& Yokoi}{Nomoto
  et~al.}{1984}]{NTY84}
Nomoto K.,  Thielemann F.-K.,    Yokoi K.,  1984, ApJ, 286, 644

\bibitem[\protect\citeauthoryear{Peletier, Vazdekis, Arribas, del Burgo,
  Garc{\'i}a-Lorenzo, Guti{\'e}rrez, Mediavilla \& Prada}{Peletier
  et~al.}{1999}]{Peletal99}
Peletier R.~F.,  Vazdekis A.,  Arribas S.,  del Burgo C.,  Garc{\'i}a-Lorenzo
  B.,  Guti{\'e}rrez C.,  Mediavilla E.,    Prada F.,  1999, MNRAS, 310, 863

\bibitem[\protect\citeauthoryear{Puzia, Saglia, Kissler-Patig, Maraston,
  Greggio, Renzini \& Ortolani}{Puzia et~al.}{2002}]{Puzetal02}
Puzia T.,  Saglia R.~P.,  Kissler-Patig M.,  Maraston C.,  Greggio L.,  Renzini
  A.,    Ortolani S.,  2002, A\&A, 395, 45

\bibitem[\protect\citeauthoryear{Saglia, Maraston, Thomas, Bender \&
  Colless}{Saglia et~al.}{2002}]{Sagetal02}
Saglia R.~P.,  Maraston C.,  Thomas D.,  Bender R.,    Colless M.,  2002, ApJ,
  579, L13

\bibitem[\protect\citeauthoryear{{Shetrone}, {Venn}, {Tolstoy}, {Primas},
  {Hill} \& {Kaufer}}{{Shetrone} et~al.}{2003}]{Sheetal03}
{Shetrone} M.,  {Venn} K.~A.,  {Tolstoy} E.,  {Primas} F.,  {Hill} V.,
  {Kaufer} A.,  2003, AJ, 125, 684

\bibitem[\protect\citeauthoryear{Shetrone, C{\^o}t{\'e} \& Sargent}{Shetrone
  et~al.}{2001}]{SCS01}
Shetrone M.~D.,  C{\^o}t{\'e} P.,    Sargent W. L.~W.,  2001, ApJ, 548, 592

\bibitem[\protect\citeauthoryear{Terlevich \& Forbes}{Terlevich \&
  Forbes}{2002}]{TF02}
Terlevich A.,  Forbes D.,  2002, MNRAS, 330, 547

\bibitem[\protect\citeauthoryear{Terlevich, Diaz \& Terlevich}{Terlevich
  et~al.}{1990}]{TDT90}
Terlevich E.,  Diaz A.~I.,    Terlevich R.,  1990, MNRAS, 242, 271

\bibitem[\protect\citeauthoryear{Thielemann, Nomoto \& Hashimoto}{Thielemann
  et~al.}{1996}]{TNH96}
Thielemann F.-K.,  Nomoto K.,    Hashimoto M.,  1996, ApJ, 460, 408

\bibitem[\protect\citeauthoryear{Thomas, Greggio \& Bender}{Thomas
  et~al.}{1998}]{TGB98}
Thomas D.,  Greggio L.,    Bender R.,  1998, MNRAS, 296, 119

\bibitem[\protect\citeauthoryear{Thomas, Greggio \& Bender}{Thomas
  et~al.}{1999}]{TGB99}
Thomas D.,  Greggio L.,    Bender R.,  1999, MNRAS, 302, 537

\bibitem[\protect\citeauthoryear{Thomas, Maraston \& Bender}{Thomas
  et~al.}{2002}]{TMB02b}
Thomas D.,  Maraston C.,    Bender R.,  2002, Ap\&SS, 281, 371

\bibitem[\protect\citeauthoryear{Thomas, Maraston \& Bender}{Thomas
  et~al.}{2003}]{TMB03}
Thomas D.,  Maraston C.,    Bender R.,  2003, MNRAS, 339, 897

\bibitem[\protect\citeauthoryear{Trager, Faber, Worthey \&
  Gonz{\'{a}}lez}{Trager et~al.}{2000a}]{Traetal00a}
Trager S.~C.,  Faber S.~M.,  Worthey G.,    Gonz{\'{a}}lez J.~J.,  2000a, AJ,
  119, 164

\bibitem[\protect\citeauthoryear{Trager, Faber, Worthey \&
  Gonz{\'{a}}lez}{Trager et~al.}{2000b}]{Traetal00b}
Trager S.~C.,  Faber S.~M.,  Worthey G.,    Gonz{\'{a}}lez J.~J.,  2000b, AJ,
  120, 165

\bibitem[\protect\citeauthoryear{Trager, Worthey, Faber, Burstein \&
  Gonz{\'{a}}lez}{Trager et~al.}{1998}]{Traetal98}
Trager S.~C.,  Worthey G.,  Faber S.~M.,  Burstein D.,    Gonz{\'{a}}lez J.~J.,
   1998, ApJS, 116, 1

\bibitem[\protect\citeauthoryear{Tripicco \& Bell}{Tripicco \&
  Bell}{1995}]{TB95}
Tripicco M.~J.,  Bell R.~A.,  1995, AJ, 110, 3035

\bibitem[\protect\citeauthoryear{Vazdekis, Peletier, Beckmann \&
  Casuso}{Vazdekis et~al.}{1997}]{Vazetal97}
Vazdekis A.,  Peletier R.~F.,  Beckmann J.~E.,    Casuso E.,  1997, ApJS, 111,
  203

\bibitem[\protect\citeauthoryear{Whitmore, McElroy \& Tonry}{Whitmore
  et~al.}{1985}]{WMT85}
Whitmore B.~C.,  McElroy D.~B.,    Tonry J.~L.,  1985, ApJS, 59, 1

\bibitem[\protect\citeauthoryear{Woosley \& Weaver}{Woosley \&
  Weaver}{1995}]{WW95}
Woosley S.~E.,  Weaver T.~A.,  1995, ApJS, 101, 181

\bibitem[\protect\citeauthoryear{Worthey}{Worthey}{1992}]{Worthey92}
Worthey G.,  1992, Phd~thesis, University of California, Santa Cruz

\bibitem[\protect\citeauthoryear{Worthey, Faber \& Gonz{\'{a}}lez}{Worthey
  et~al.}{1992}]{WFG92}
Worthey G.,  Faber S.~M.,    Gonz{\'{a}}lez J.~J.,  1992, ApJ, 398, 69

\bibitem[\protect\citeauthoryear{Worthey}{Worthey}{1998}]{Worthey98}
Worthey G.,  1998, PASP, 110, 888

\end{thebibliography}
\end{document}